\documentclass[11pt,a4paper]{article}
\usepackage[T1]{fontenc}
\usepackage[utf8]{inputenc}
\usepackage{lmodern}
\usepackage{amsmath,amssymb}
\usepackage{graphicx}
\usepackage{capt-of}
\usepackage{textcomp}
\usepackage[hidelinks]{hyperref}
\usepackage[margin=1in]{geometry}
\usepackage{xurl}
\usepackage[hidelinks]{hyperref}

\usepackage[
  backend=biber,
  style=numeric,
  sorting=none,
  giveninits=true,
  maxbibnames=999
]{biblatex}
\DeclareNameAlias{default}{family-given}
\DeclareNameAlias{sortname}{family-given}

\begin{document}

\title{Generative modelling for multivariate spatio-temporal downscaling of monthly climate fields}
\author{Yu Huang\textsuperscript{1,2}, Shangshang Yang\textsuperscript{1,2},
Sebastian Bathiany\textsuperscript{1,2}, Michael Aich\textsuperscript{1,2},\\
Philipp Hess\textsuperscript{1,2}, Niklas Boers\textsuperscript{1,2,3}\\[1ex]
\small \textsuperscript{1}Munich Climate Center and Earth System Modelling Group,\\
\small Department of Aerospace and Geodesy, School of Engineering and Design,\\
\small Technical University of Munich, Germany\\
\small \textsuperscript{2}Complexity Science, Potsdam Institute for Climate Impact Research, Germany\\
\small \textsuperscript{3}Department of Mathematics and Statistics, University of Exeter, UK\\[1ex]
\small Correspondence: Yu Huang (\href{mailto:y.huang@tum.de}{y.huang@tum.de})
and Shangshang Yang (\href{mailto:shangshang.yang@tum.de}{shangshang.yang@tum.de})}
\date{}
\maketitle

\begin{abstract}

Inferring climate fields at higher spatio-temporal resolutions from coarse representations is essential for accessing socioeconomic and environmental impacts. The downscaling of modelled climate fields faces a trade-off between data storage constraints, simulation efficiency, and resolution requirements. While many methods for spatial downscaling have been proposed, joint spatio-temporal downscaling has received comparable little attention. When deriving, for example, daily fields from climate datasets that are available only as monthly averages at coarser spatial resolution, it is highly challenging to faithfully reproduce extreme event statistics and spatiotemporal consistency. To address this task, we introduce a generative machine learning approach, using a diffusion model to jointly downscale monthly temperature and precipitation fields into temporally consistent daily fields at higher spatial resolution. We demonstrate that our approach can generate high-dimensional multivariate climate fields at daily temporal and 0.25\textdegree{} spatial resolution from monthly input fields at 1\textdegree{} spatial resolution, exhibiting coherent temporal dynamics and physically plausible spatial structures. In particular, by incorporating a conditional guidance technique, we show that a pre-trained diffusion model can be directly leveraged to efficiently generate daily high-resolution fields conditioned on monthly inputs, without requiring additional fine-tuning or retraining. Our probabilistic approach thus provides an efficient and generalizable framework for spatio-temporal climate downscaling, enabling flexible climate fields generation for climate impact assessments and digital Earth simulator constructions.

\end{abstract}


\section{Introduction}  
Accurate representation of climate variables at sufficiently high spatio-temporal scales is essential for understanding and quantifying the ecological and socioeconomic impacts of climate change \cite{RN23, sofaer2017designing, buster2024high, kim2023exposure, yoon2026extreme, heubel2025forest, he2025rainfall,hess2025fast}. Storage or computation constraints often require climate datasets to be archived at coarse spatial and temporal resolutions, limiting their direct applicability in climate impact assessment studies. Moreover, several digital Earth initiatives \cite{hoffmann2023destination,li2023big,schicker2025beyond,cui2024digital} are expected to employ online simulation approaches, in which core datasets are stored at relatively coarse spatial and temporal resolutions to reduce storage costs, while higher-resolution outputs are generated on demand for user analysis and applications. Furthermore, recently developed climate emulators enable very rapid simulations at relatively coarse spatio-temporal scales while still capturing the long-term response of climate fields to external forcings \cite{lucarini2017predicting, aengenheyster2018point, lucarini2023theoretical, womack2025rapid, sandstad2025meteorv1}. However, they typically provide only annually or monthly mean outputs. A natural avenue for improving fast climate emulators is the development of efficient spatio-temporal downscaling methods, motivated by the practical need to produce fine-grained outputs from coarse-resolution data. Bridging such scale gap requires efficient and generalizable downscaling methods, yet remains challenging due to complex spatiotemporal dependencies and the difficulty to consistently preserve daily extreme event behavior that averages out at coarse resolutions \cite{hasegawa2021extreme, miller2025statistical,hernanz2022evaluation}.

Existing downscaling approaches for climate data generally fall into three categories: dynamical, statistical, and machine learning--based methods. Dynamical downscaling \cite{giorgi2019thirty, luo2022dynamic}, which employs regional climate models driven by large-scale boundary conditions, produces physically consistent high-resolution outputs, but is computationally intensive. Moreover, such numerical simulations require a full set of climate variables, rather than a single variable of interest. Statistical downscaling methods \cite{tabari2021comparison, Maraun2010, maraun2018statistical, hernanz2022evaluation, palmer2019stochastic}, such as interpolation-based methods, regression-based models and stochastic weather generators, are computationally efficient but may misrepresent high-dimensional spatiotemporal structures and extreme values \cite{miller2025statistical,hernanz2022evaluation,hess2025fast}, particularly for highly non-linear and intermittent variables such as precipitation. More recently, machine learning (ML)--based approaches have shown remarkable success in learning complex dependencies and patterns from data \cite{bi2022pangu, lam2023learning, price2025probabilistic, yang2025generative, huang2020reconstructing}, particularly for high-dimensional spatiotemporal climate fields. Recent advances in probabilistic generative machine learning (GML) have further demonstrated strong effectiveness in producing realistic climate fields, effectively capturing heavy-tailed probability distributions, intermittency, and spatial heterogeneity of variables such as precipitation \cite{aich2026conditional, hess2022physically, hess2025fast, hess2025generating}. Early GML approaches, such as variational autoencoders, often struggle to generate high-quality spatial precipitation fields. This limitation arises from their maximum-likelihood--based training objective, which tends to prioritize average reconstructions and consequently leads to blurred outputs and a loss of fine-scale spatial detail \cite{pan2023probabilistic, szwarcman2024quantizing}. Generative adversarial networks are able to generate high-resolution fields but suffer from training instability and are prone to mode collapse \cite{dhariwal2021diffusion}. Diffusion models (DMs) are state-of-the-art generative machine learning frameworks, which are inspired by physical diffusion processes. They allow for generating high-resolution and diverse images while maintaining stable training and inference procedures \cite{saharia2022palette,ho2022video,croitoru2023diffusion,price2025probabilistic}. Based on the denoising procedure of a reversed stochastic differential equation, DMs can be trained to generate climate fields with high spatial resolution \cite{hess2025fast, aich2026conditional}. Newly developed GML frameworks such as Flow Matching\cite{lipman2022flow}, have also shown promising performance comparable to DMs, whereas DMs currently benefit from a larger user base and more mature developments in conditional generation techniques.

In contrast to ML tasks in weather forecasting, which aim to produce deterministic or probabilistic forecasts that can be directly paired with the ground truth, downscaling climate fields lacks paired high-resolution ground truth data. In addition, the scarcity of high-resolution observations, combined with the inability of climate models to fully replicate historical climate due to the chaotic behavior of geophysical fluid dynamics, renders downscaled simulations inherently probabilistic \cite{palmer2019stochastic, giorgi2019thirty, luo2022dynamic, hess2025fast}. Consequently, downscaling emphasizes maintaining consistency in spatiotemporal distributions and accurately representing extreme values. Some studies have recently employed GML methods for spatial downscaling of climate fields. High-resolution datasets, typically ERA5 \cite{hersbach2020era5} or higher-resolution climate simulation data, are used to train ML models, such as DMs, to obtain a generator capable of generating high-resolution climate fields. The pretrained DM can then be applied for spatial downscaling using conditional guidance techniques \cite{meng2021sdedit, ho2022classifier, aich2026conditional, hess2025fast, yang2025generative, aich2026wind}. 

In addition to spatial downscaling \cite{aich2026conditional, hess2025fast, lopez2025dynamical, rampal2025reliable,tomasi2025can}, GML frameworks have also demonstrated skill in temporal downscaling \cite{bassetti2024diffesm, aich2026wind}, as well as joint spatio-temporal downscaling \cite{glawion2025global,schmidt2025generative}. Generative machine learning for processing climate data is a rapidly evolving area. For instance, Glawion et al. \cite{glawion2025global} employed GAN to downscale precipitation from hourly to sub-hourly. Schmidt et al. \cite{schmidt2025generative} leveraged DMs to downscale 6-hourly temperature, winds and pressure to their hourly representations. In this work we focus on temporal downscaling from monthly to daily representations, which is particularly crucial for socioeconomic and ecological simulations. Bassetti et al. \cite{bassetti2024diffesm} presented a method for temporally downscaling monthly Earth system model (ESM) data to daily resolution. They input the monthly map as the prior condition into the DM, and the corresponding 30 consecutive daily maps within this month are set as the inference target. Trained on the extensive samples of climate fields provided by the ESM simulation, their temporal downscaling model can effectively generate daily outputs from monthly data under different global warming scenarios. However, their pre-trained DM, developed using relatively coarse-resolution ESM data, only performs temporal downscaling, and cannot be directly applied to high-resolution observational climate data. Retraining this diffusion model (DM) with high-resolution observational data is further constrained by GPU memory limitations, as loading 30-day sequences of global high-resolution temperature and precipitation fields can exceed memory capacity. To reduce GPU memory requirements, in this study we consider using an autoregressive approach \cite{price2025probabilistic,hess2025generating} to model temporal consistency. In addition, we aim to achieve both temporal and spatial downscaling within a single diffusion model. 

In this work, we introduce a monthly-fields-guided generative diffusion model, which leverages a pre-trained diffusion model to perform joint spatio-temporal downscaling. Conditioning on monthly fields via conditional generation techniques \cite{song2020score, ho2022classifier, chung2025diffusion, hess2025fast}, the model generates temporally consistent daily fields in an autoregressive manner. This design enables the model to preserve temporal dynamics while flexibly representing uncertainty at fine spatio-temporal scales. To demonstrate our approach, we use daily fields at 0.25\textdegree{} resolution, as well as monthly fields at 1\textdegree{} resolution, from the widely used ERA5 reanalysis~\cite{hersbach2020era5,chai2022evaluation,lavers2022evaluation,bandhauer2022evaluation,tarek2020evaluation}. Temperature and precipitation, which are central to climate impact analyses, are selected for joint multivariate spatio-temporal downscaling. 
Our approach is capable of generating high-dimensional multivariate climate fields at fine spatio-temporal scales, translating coarse-resolution monthly inputs into daily fields at a global spatial resolution of $0.25^\circ$. The downscaled outputs exhibit power spectra consistent with reference data at small scales, while faithfully preserving spatio-temporal coherence and temporal dynamics. Our diffusion model also captures extreme-value behaviour and inter-variable covariance. Together, these results demonstrate a skillful and efficient framework for generating fine-grained climate data, with potential applications in socioeconomic and ecological impact assessments.

In Section 2, we provide a detailed description of the data preprocessing and our methods. In Section 3, we present the generated daily fields and evaluate their key statistical properties. Finally, in Section 4, we summarize our work and discuss potential directions for further improving our spatiotemporal downscaling method.

\section{Methodology}
\subsection{Data}
We use daily 2\,m surface air temperature (\text{tas}, in units of $^{\circ}$C) and total daily precipitation (\text{pr}, in units of mm/day) from the ERA5 reanalysis for the period 1979--2023 to train our diffusion model. The original ERA5 data are provided at an hourly temporal resolution with a spatial resolution of $0.25^{\circ} \times 0.25^{\circ}$. These data are temporally aggregated to daily values prior to model training. 

We also employ the monthly averaged ERA5 reanalysis fields. The monthly data are spatially downsampled to a coarser resolution of $1^{\circ} \times 1^{\circ}$, acting as a testbed for the  performance of our approach for spatio-temporal downscaling. 

During data preprocessing, precipitation $\text{pr}$ is initially transformed using $\log_{10}(\text{pr} + 1)$. Subsequently, for each spatial field, either temperature or transformed precipitation, we apply consistent normalization coefficients to linearly rescale the values to the range $[-1,1]$. 

The ERA5 data from 1979 to 2014 is randomly partitioned into a training set (90\%) and a validation set (10\%), while data from 2015 to 2023 is reserved for evaluation. At inference, the DM can jointly generate daily temperature and precipitation fields, given the daily fields from the preceding two days, along with the corresponding monthly fields.

\subsection{Diffusion model specification and configurations}

Diffusion models (DMs) are generative machine learning models that excel at learning probability distributions of high-dimensional data and generating high-resolution natural images and videos from noise. In this work, we employ a DM to model the joint probability distribution of daily and monthly temperature and precipitation fields. This generative modeling framework is designed to enable the controllable generation with a pretrained diffusion model through the incorporation of posterior-guidance approaches \cite{song2020score, chung2025diffusion, ho2022classifier, hess2025fast}. 

The global fields of temperature and precipitation on a given day (\textit{d}) are represented by the matrix $\mathbf{X}_d$, while their temporal average (30 days) is represented by the matrix $\mathbf{X}_m$. We define the combined state $\mathbf{X} := (\mathbf{X}_d, \mathbf{X}_m)$, where $\mathbf{X} \in \mathbb{R}^{n}$ and $n$ denotes the data dimension. Correspondingly, the initial matrices for diffusion model inference, denoted as $\mathbf{Z} \in \mathbb{R}^{n}$, are initialized with values sampled from a Gaussian distribution $\mathcal{N}(0, I)$ ($I \in \mathbb{R}^{n}$). In addition, the daily temperature and precipitation fields from one and two days (\textit{d}-1 and \textit{d}-2) prior to the target day (\textit{d}) are incorporated as the matrices $\mathbf{X}_{d-1}$ and $\mathbf{X}_{d-2}$, respectively. Then the learning objective of the DM is formulated as estimating the following conditional probability distribution:
\begin{equation}
    P(\mathbf{X}_d, \mathbf{X}_m \mid \mathbf{Z}, \mathbf{X}_{d-1}, \mathbf{X}_{d-2}).
\end{equation}
Initialized from a Gaussian distribution, the DM generates daily and monthly fields jointly, conditioned on the daily states of the two preceding days. The model specification and configuration are described below. 

Following the standard denoising diffusion probabilistic model (DDPM) framework~\cite{ho2020denoising, song2020score}, training proceeds via a forward diffusion process (Fig.~1a), in which Gaussian noise is progressively injected into the target fields according to a stochastic differential equation (SDE): 
\begin{equation}
\  d\mathbf{Z}_{t}
= f(\mathbf{Z}_{t},t)dt + g(t)dW, \quad t: 0 \to T.
\end{equation}
Here $t\in[0,T]$ denotes the integration step of the SDE, and $W$ is the standard Wiener process. $\mathbf{Z}_{t}$ represents the intermediate state in the iteration (Fig. 1a). The drift term $f(\mathbf{Z}_t,t)$ and diffusion term $g(t)$ are predefined according to the configuration of score-based generative modeling \cite{song2020score}.

The denoising inference of the DM follows the numerical solution of the associated reverse-time SDE, using the score function learned via score matching \cite{song2020score,yang2025generative}. The reversed SDE for denoising and generation, evolving $t$ from $T$ to $0$ (reverse for the SDE), is given by:

\begin{equation}
d\mathbf{Z}_t
=
\Bigg[
\mu(\mathbf{Z}_t,t)
- g^{2}(t) \nabla_{\mathbf{Z}_t} \log p_{t}
(\mathbf{Z}_t \mid \mathbf{X}_{d-1}, \mathbf{X}_{d-2})
\Bigg]
dt
+ g(t) d\overline{W}, \quad t: T \to 0.
\end{equation}

Since the score function $\nabla_{\mathbf{Z}_{t}}\log p_{t}(\mathbf{Z}_{t} \mid \mathbf{X}_{d-1}, \mathbf{X}_{d-2})$ is not analytically tractable, the DM framework explicitly approximates it with a neural network: $s_{\theta}(\mathbf{Z}_{t}, t, \mathbf{X}_{d-1}, \mathbf{X}_{d-2}) \approx \nabla_{\mathbf{Z}_{t}}\log p_{t}(\mathbf{Z}_{t} \mid \mathbf{X}_{d-1}, \mathbf{X}_{d-2})$.
 Each integration step in Eq. (3) is conditioned on $\mathbf{X}_{d-1}$ and $\mathbf{X}_{d-2}$ (hereafter referred to as prior daily condition), and it is designed to converge toward $\mathbf{X}$, which constitutes the denoising inference target.

We parameterize $s_{\theta}$ using a U-Net architecture~\cite{song2020score, liu2022feichtenhofer,yang2025generative} and train it via score matching with gradient descent~\cite{song2020score}. To improve convergence speed and sample fidelity, we adopt v-prediction during training~\cite{karras2022elucidating}. The network input comprises the noised fields $\mathbf{Z}$, concatenated along the channel dimension to form a four-channel tensor in PyTorch. Conditioning on the preceding two days is achieved by additionally concatenating $\mathbf{X}_{d-1}$ and $\mathbf{X}_{d-2}$ along the channel dimension.

We encode the diffusion timestep $t$ by adding sinusoidal positional embeddings to the model input~\cite{ho2020denoising}. The daily precipitation target is defined as the anomaly relative to the corresponding monthly precipitation, promoting consistency between the jointly generated daily and monthly fields. After inference, the anomaly is added back to the monthly field to recover the daily precipitation. Other variables (that is, monthly precipitation, monthly temperature and daily temperature) are represented in their original form rather than as anomalies.

The training of neural networks was conducted for 500 epochs using the Adam optimizer with a learning rate of $1\times10^{-4}$ and a batch size of 1. Training was conducted on an NVIDIA H100 GPU. The reverse-time stochastic differential equation (SDE) was solved numerically from $t=T$ to $t=0$ via the Euler--Maruyama scheme with 2000 time discretization steps \cite{ho2020denoising,song2020score}. During the inference stage, computational efficiency is improved by adopting a second-order probability flow ordinary differential equation solver \cite{song2020denoising,yang2025generative,aich2026conditional} and reducing the integration steps to 200. This strategy substantially accelerates sampling while retaining high-quality generation performance.

\begin{center}
\includegraphics[width=0.85\linewidth,keepaspectratio]{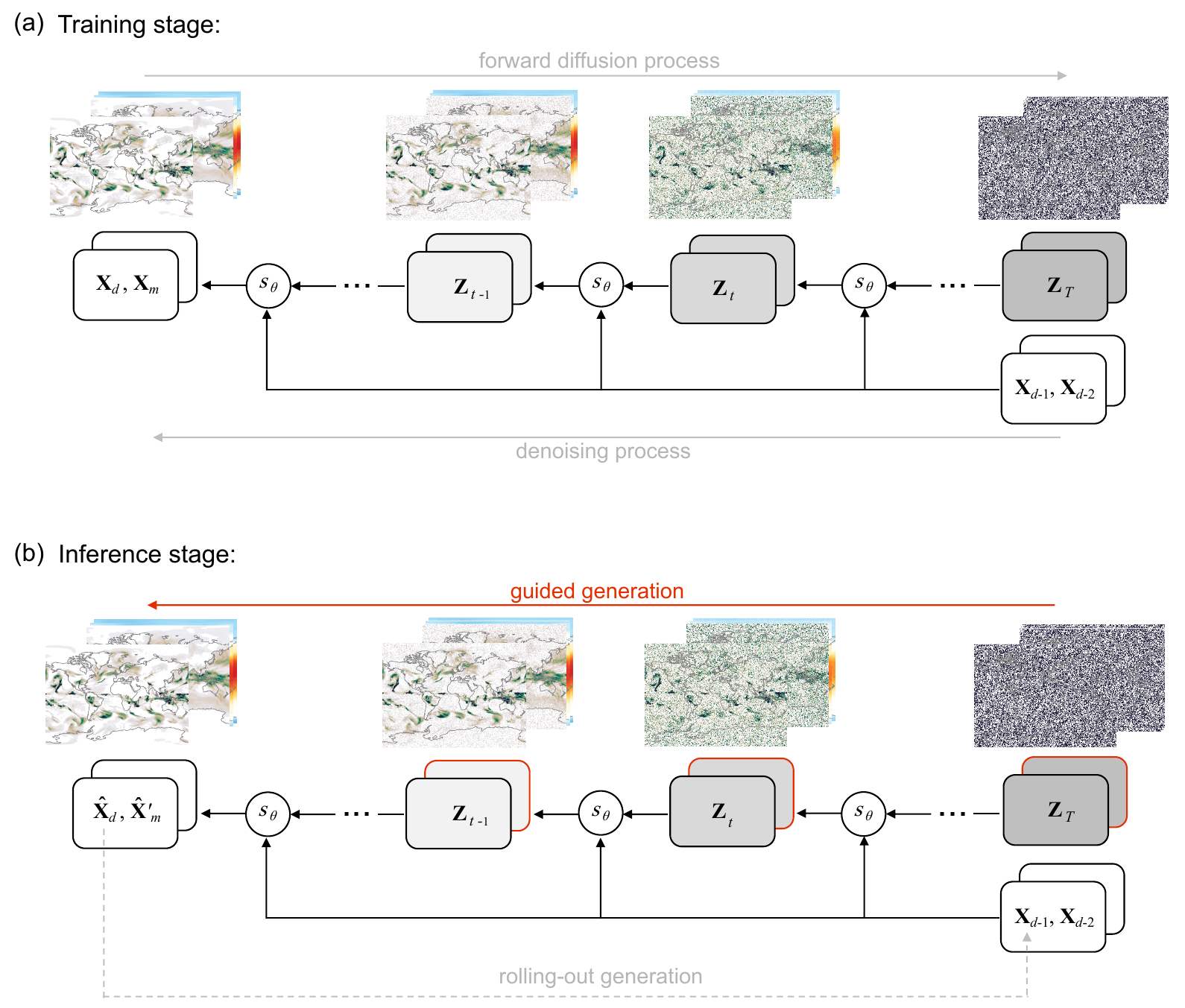}
\end{center}
\captionof{figure}{\textbf{Illustration of the monthly-fields-informed diffusion generator.} 
(a) During training, the selected climate fields are represented as multi-channel images, consisting of monthly ($\mathbf{X}_{m}$) and daily ($\mathbf{X}_{d}$) temperature and precipitation. They are progressively corrupted by noise through a forward diffusion process governed by a stochastic differential equation (SDE), resulting in a full noise state $\mathbf{Z}_{T}$. The backward denoising process then aims to recover the noise-free image. Initializing from a full noise sample $\mathbf{Z}_{T}$, the reversed SDE progressively denoises $\mathbf{Z}_{t}$ to infer $\mathbf{Z}_{t-1}$. The joint temperature and precipitation fields from the previous two days ($\mathbf{X}_{d-1}$ and $\mathbf{X}_{d-2}$) serve as condition during each iterative state of $\mathbf{Z}_{t}$. After training, the diffusion model here can generate $\mathbf{X}_{m}$ and $\mathbf{X}_{d}$ jointly by initializing with pure noise $\mathbf{Z}_{T}$. (b) During inference, the known monthly fields of temperature and precipitation $\mathbf{X}^{'}_{m}$ are used to guide the generation of the pretrained diffusion model. Our model operates in an autoregressive setting where the output $\mathbf{X}^{'}_{d}$ is used as condition for the next frame. }\label{fig1}

\subsection{Guided generation for daily fields}
The pretrained DM jointly models the daily and monthly atmospheric states. In implementation, the daily and monthly fields are concatenated along the channel dimension, equivalently forming a single ``generalized'' image with multiple channels. Under the Bayesian formulation, the known monthly fields can be incorporated as observations through the likelihood term \cite{song2020score, chung2025diffusion, meng2021sdedit} to guide the diffusion model in generating the corresponding daily fields.

To match the input dimensionality required by the pretrained DM, the aforementioned monthly ERA5 data with a spatial resolution of $1^{\circ} \times 1^{\circ}$ are mapped to a resolution of $0.25^{\circ} \times 0.25^{\circ}$ using bilinear spatial interpolation for both precipitation and temperature fields. Subsequently, linear temporal interpolation \cite{maraun2018statistical,lim2025climforge} is applied to the monthly time series at each grid cell to match the dimension in the pretrained DM. The spatio-temporal linear interpolation employed here is a commonly used statistical downscaling approach \cite{maraun2018statistical,lim2025climforge}. We adopt it as a preprocessing step prior to integrating the monthly data with the pretrained DM. It helps to avoid discontinuities arising from the coarse resolution of the original data and ensures consistency with the dimensional requirements of the pretrained DM.

At the inference stage, we apply the linearly conditional guidance approach \cite{song2020score, chung2025diffusion} to guide the generation process of the pretrained diffusion model with the known monthly fields $\mathbf{X}'_m$ (Supplementary Note 1). Accordingly, Eq.~(3) is further formulated as the conditional generation process:
\begin{equation}
d\mathbf{Z}_{t}
=
\Bigg[
\mu(\mathbf{Z}_{t},t)
- g^{2}(t)
\left(
s_{\theta}(\mathbf{Z}_{t}, t, \mathbf{X}_{d-1}, \mathbf{X}_{d-2})
+
s_{g}(\mathbf{Z}_{t}, \mathbf{X}'_m)
\right)
\Bigg]
dt
+
g(t) d\overline{W},
\end{equation}
where $s_{\theta}$ denotes the neural-networks-approximated score function within the pretrained diffusion model, and $s_{g}(\mathbf{Z}_{t}, \mathbf{X}'_m)$ is the posterior-guided
conditional score function, predefined in the linearly conditional approach \cite{song2020score, chung2025diffusion} (See Supplementary Note 1 for details). During the guided generation process (Fig.~1b), the preprocessed monthly fields are used to guide the diffusion model. After generating the daily temperature and precipitation fields for a given day, the output is fed back to update the model's prior daily conditions (Fig.~1b), thereby enforcing a rolling-day generation scheme. The resulting monthly-fields-informed daily field generator thus functions as a spatio-temporal downscaling model. 
To initialize the model at the start of each simulation, daily fields from two consecutive days in historical data with matching calendar dates are used as the initial prior conditions. The model is then run for 30 daily steps as a spin-up period to remove initial transients and achieve a stationary state \cite{washington2005climate, strogatz2014nonlinear}.

\section{Results}

\subsection{Generating daily fields from monthly data}
Figure 2 presents three consecutive daily fields generated by the diffusion model. Compared with the original monthly fields (Fig. 2a), the generated daily fields provide a more detailed representation of weather events such as cold waves, heat waves, and heavy precipitation. The spatial structure and temporal evolution of the associated large-scale weather systems is evident across three consecutive days (Fig. 2b).

\begin{center}
\includegraphics[width=0.7\textwidth,keepaspectratio]{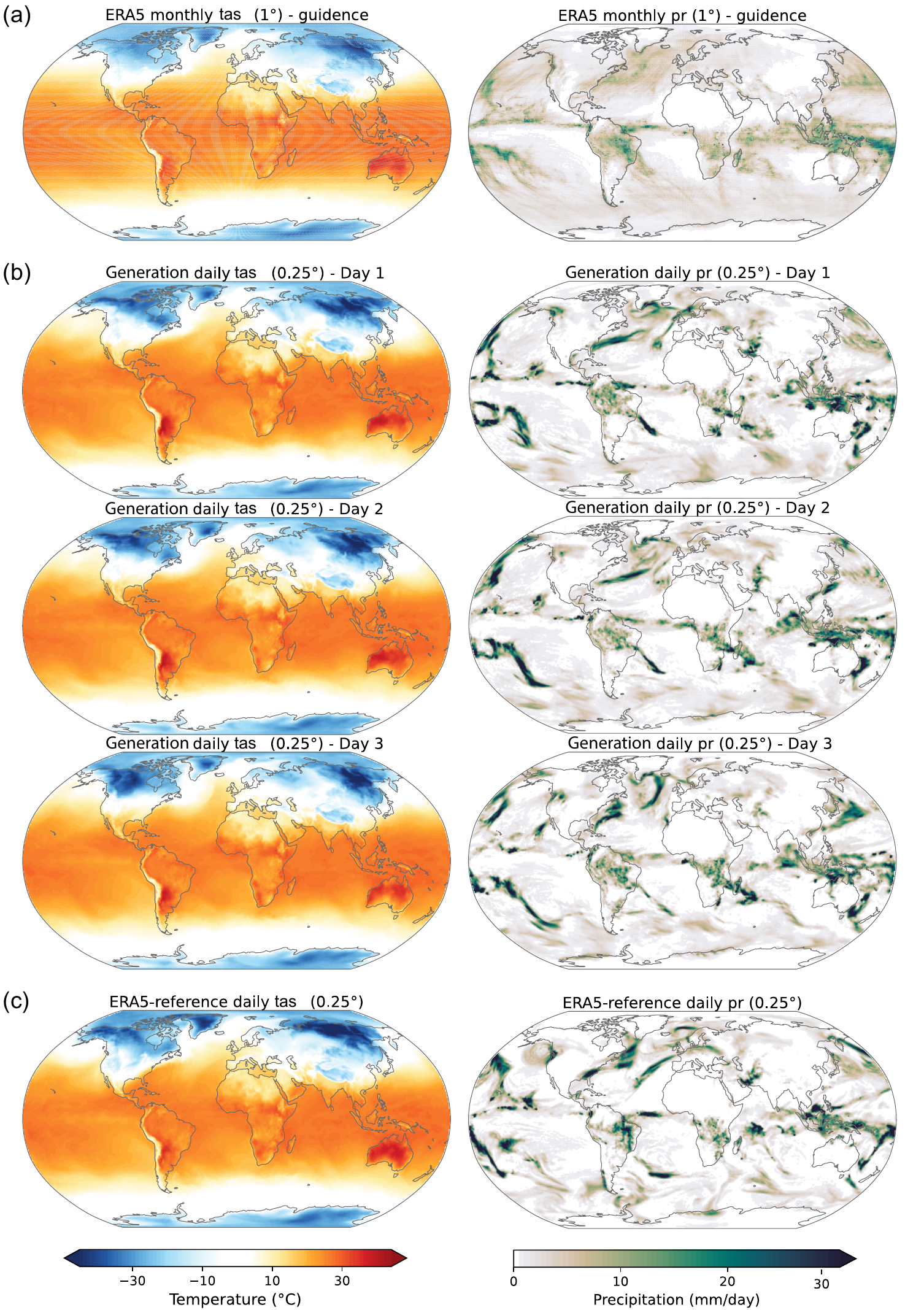}
\end{center}
\captionof{figure}{\textbf{Generated daily fields by spatio--temporal downscaling of monthly fields.} (a) The monthly temperature (tas) and precipitation (pr) fields in January 2015 are shown, with a resampled spatial resolution of $1^{\circ}$. (b) After spatio-temporal downscaling by the diffusion model, three of the generated consecutive daily fields with the enhanced spatial resolution of $0.25^{\circ}$ are visualized. (c) The referenced ERA5 daily fields from the same month are illustrated.}\label{fig2}\par\vspace{0.3cm}

Hovmöller diagrams in Fig. 3a further illustrate the 180-day evolution of temperature and precipitation in the Northern mid-latitudes. Within this latitude band, generated daily high-resolution temperature and precipitation fields both exhibit the correct eastward propagating wave-like pattern \cite{li2024key, wang2025evidence}. Moreover, the wintertime surface air temperatures over land are generally lower than over the ocean, while precipitation is stronger over ocean regions. The generated daily fields closely replicate these spatiotemporal patterns observed in the ERA5 reference data. Similar Hovmöller diagrams for the tropics (Fig. 3b) and the Southern mid-latitudes (Fig. 3c) show that the generated fields likewise capture the spatio-temporal patterns seen in ERA5. In contrast to machine-learning-based weather forecasting, the weather evolution in the downscaled output of diffuion model cannot be expected to match the ground truth daily fields for the same month (Fig. 2c), but rather achieves consistency in its summary statistical properties (see following sections). The generated daily fields closely resemble the daily ERA5 data in terms of spatial distribution characteristics, such as the relative arrangement of high- and low-value regions for temperature and precipitation (Figs. 2b and 2c). This similarity stems from their shared dependence on the same large-scale circulation structures within the background monthly fields.

\begin{center}
\includegraphics[width=0.8\textwidth,keepaspectratio]{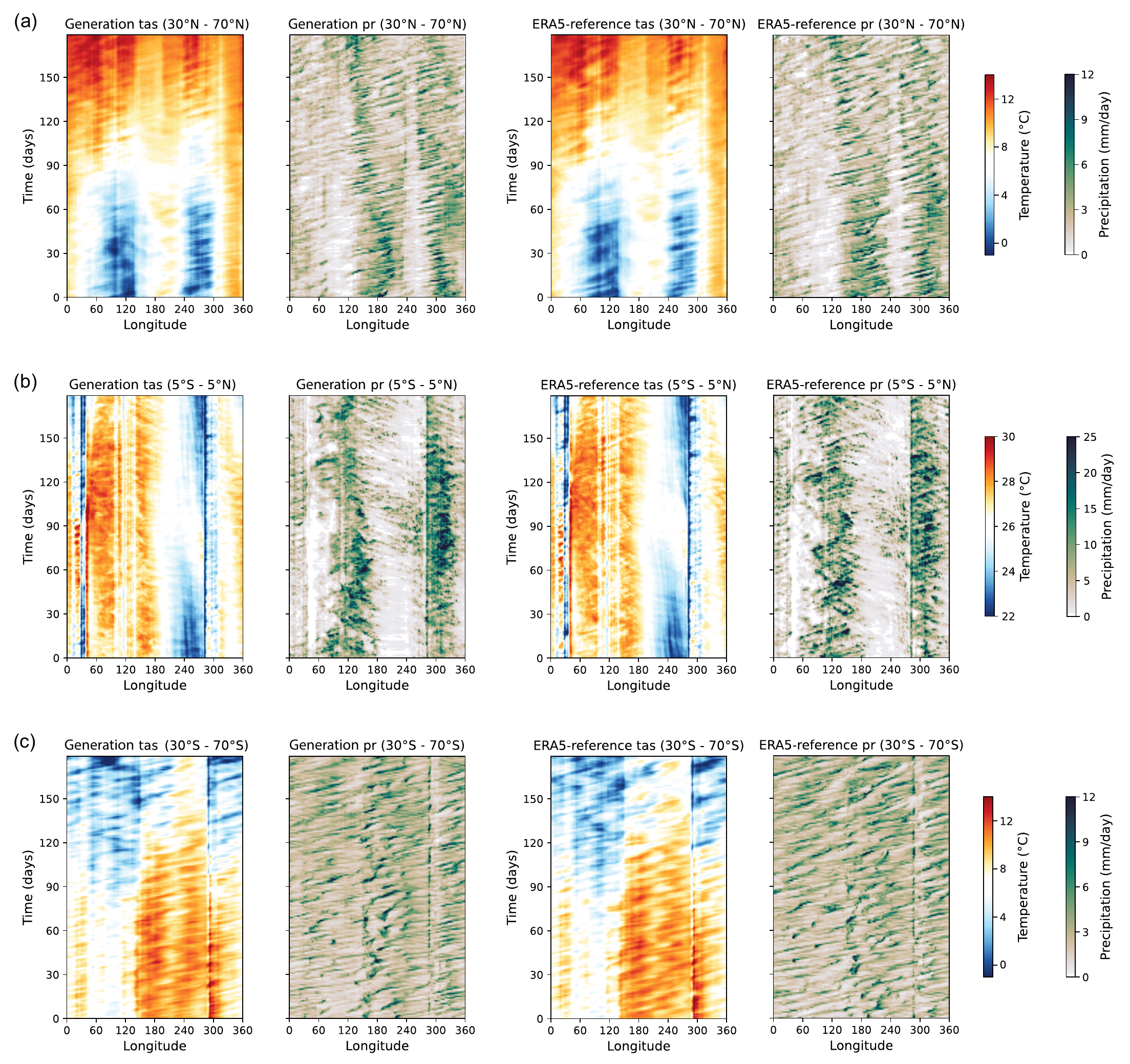}
\end{center}
\captionof{figure}{\textbf{Spatio-temporally evolutionary patterns of generated temperature and precipitation.} The time period from 2015-01-01 to 2015-6-30 is taken as the example to illustrate the evolution of the daily fields. (a) Temperature (tas) and precipitation (pr) within latitude range from $30^{\circ}$N to $70^{\circ}$N are spatially averaged across latitudes, respectively. The resulting temperature and precipitation fields are then treated as functions of time and longitude. The results for the diffusion-model generation and ERA5 reference datasets are shown on the left and right, respectively. (b) Same as (a), but for latitude range from $5^{\circ}$S to $5^{\circ}$N. (c) Same as (a), but for latitude range from $30^{\circ}$S to $70^{\circ}$S. }\label{fig3}\par\vspace{0.3cm}

We subsequently use monthly fields from 2015--2023 to condition the diffusion model and generate continuous daily outputs. Figs. 4a--d compare the global mean time series of the generated daily fields with the ERA5 reference daily fields, demonstrating that the generated data preserve the long-term temporal trends of the ERA5 dataset.

The monthly means of the generated temperature fields closely match those of the ERA5 reference (Fig. 4e). For precipitation, the monthly means of the generated fields are also highly consistent with ERA5 (Fig. 4f), although slight discrepancies remain and the curves do not fully overlap (see below for further discussion). In addition, the temporally averaged values at individual grid cells derived from the generated daily fields are in good agreement with ERA5 (Figs. 4h--j). Overall, the diffusion model demonstrates high fidelity in reproducing the mean statistical properties of both temperature and precipitation relative to the ERA5 reference.

\begin{center}
\includegraphics[width=0.65\textwidth,keepaspectratio]{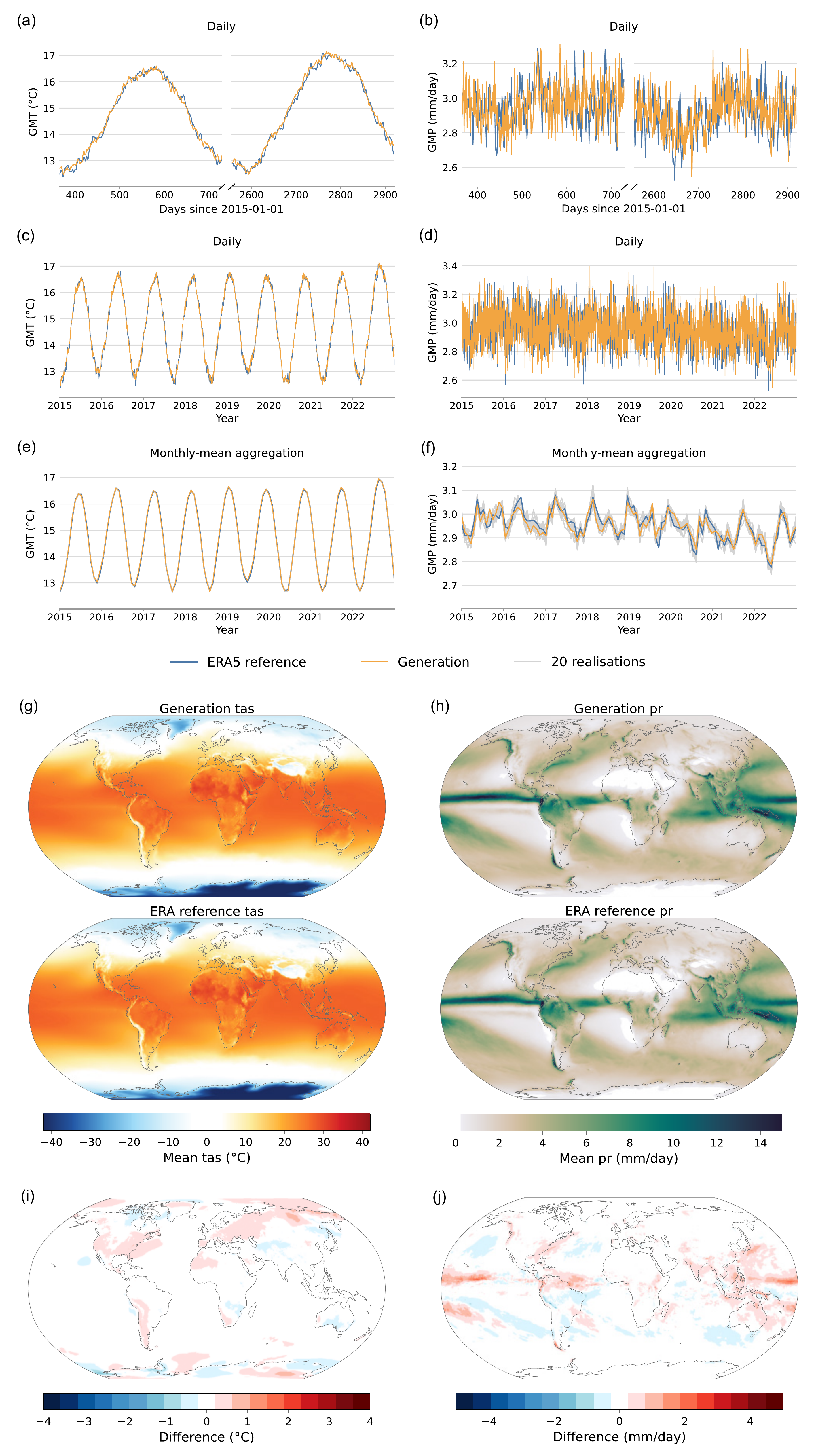}
\end{center}

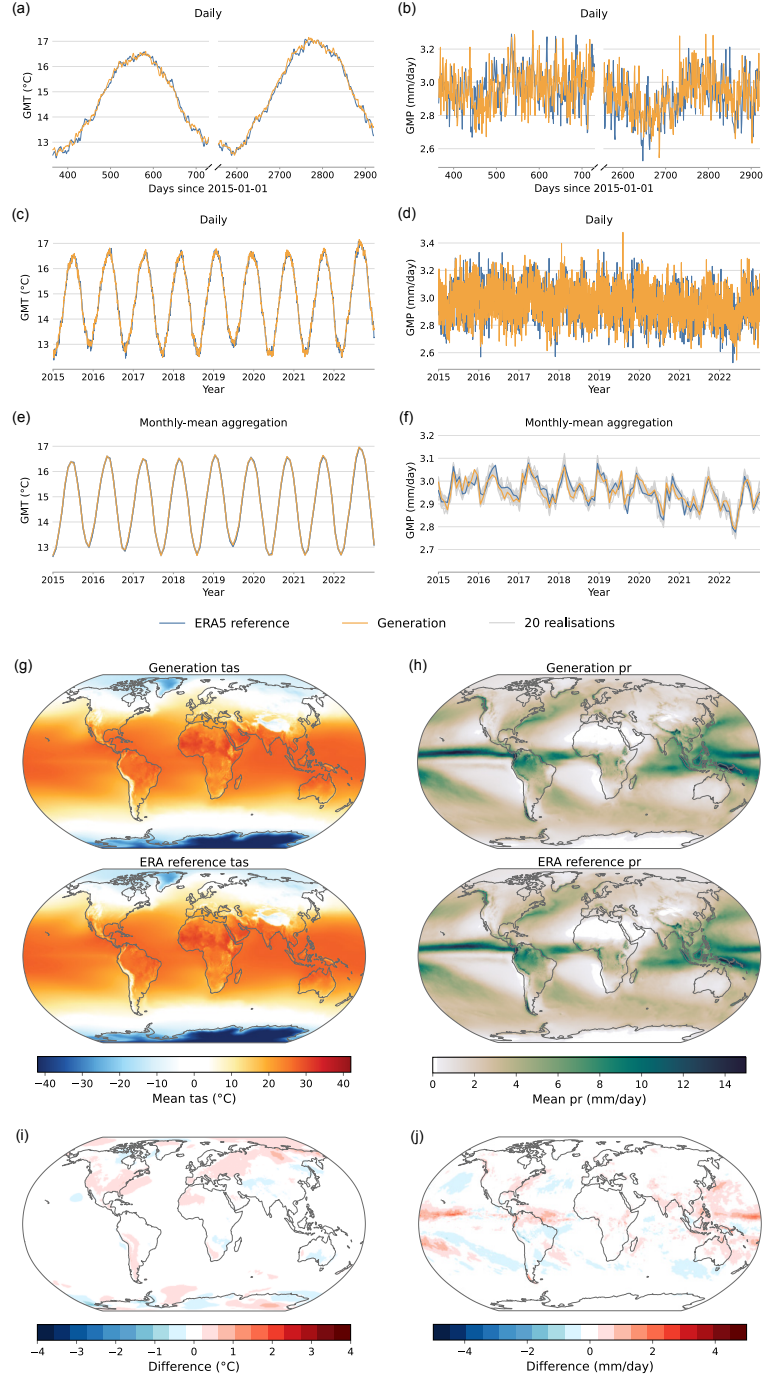
\captionof{figure}{\textbf{Comparison of temperature and precipitation between downscaling-generated fields and ERA5 reference data.} (a-b) Daily time series of global mean aggregated temperature (GMT) and precipitation (GMP) of downscaling-generated fields and ERA5 reference. Two representative one-year segments are shown for visualization clarity. (c-d) Same as (a-b), but showing the complete time series over the 2015--2023 period. (e-g) Comparison of monthly mean aggregated from downscaling-generated temperature and precipitation fields and the ERA5 reference. Given the probabilistic nature of the diffusion model, results from 20 realizations are also shown. (g-h) Comparison of temporal mean temperature and precipitation at individual grid cells between downscaling-generated daily fields and ERA5 reference. (g) Difference of mean temperature at individual grid cells between downscaling-generated daily fields and ERA5 reference. (h) Same as (g), but for the difference of precipitation. All statistics are calculated over the testing period 2015--2023.}\label{fig4}\par\vspace{0.3cm}

We note that the diffusion model in this study is trained to estimate the joint distribution of monthly and daily fields (Eq. (1)). This constitutes a stochastic modeling framework \cite{palmer2019stochastic}, in which the model learns probabilistic structural constraints rather than deterministic one-to-one relationships \cite{hess2025fast, aich2026generating, aich2026conditional}. The probabilistic nature of this framework is further reflected in the subtle variability among the 20 generated realizations (Figs. 4e-f).
In contrast to temperature, precipitation fields are inherently characterized by strong spatiotemporal heterogeneity \cite{pendergrass2018uneven,hess2022physically}, high-dimensional complexity \cite{sun2018review}, and pronounced multiscale variability \cite{yang2019process,yang2020evaluation}. During the sparse sampling process that transformed high-resolution precipitation fields into coarse-resolution representations, part of the high-frequency information was inevitably lost. Consequently, when the diffusion model is subsequently applied to downscale the coarse-resolution fields, the missing high-frequency structures can only be reconstructed in a probabilistic manner. This further highlights the intrinsic difficulty of spatiotemporal downscaling for precipitation fields. Therefore, although the monthly mean precipitation derived from the generated fields remains close to that of the original high-resolution precipitation fields, the two do not overlap exactly (Fig. 4f).
In this study, the precipitation fields are generated using the standard training and inference pipeline of the diffusion model. In future work, depending on the requirements of specific applications, hard constraints on monthly mean values could be incorporated either into the loss function or into the final output layer of the generative model \cite{hess2022deep,hess2022physically}.

\subsection{Evaluation on generated daily fields}
In the following, the spatio-temporal downscaling products are quantitatively examined using multiple evaluation metrics, based on daily outputs generated for the period 2015--2023. Using the $0.25^{\circ} \times 0.25^{\circ}$ ERA5 daily data as reference, we show that our approach can reproduce comparable summary statistical properties.

\begin{center}
\includegraphics[width=0.7\textwidth,keepaspectratio]{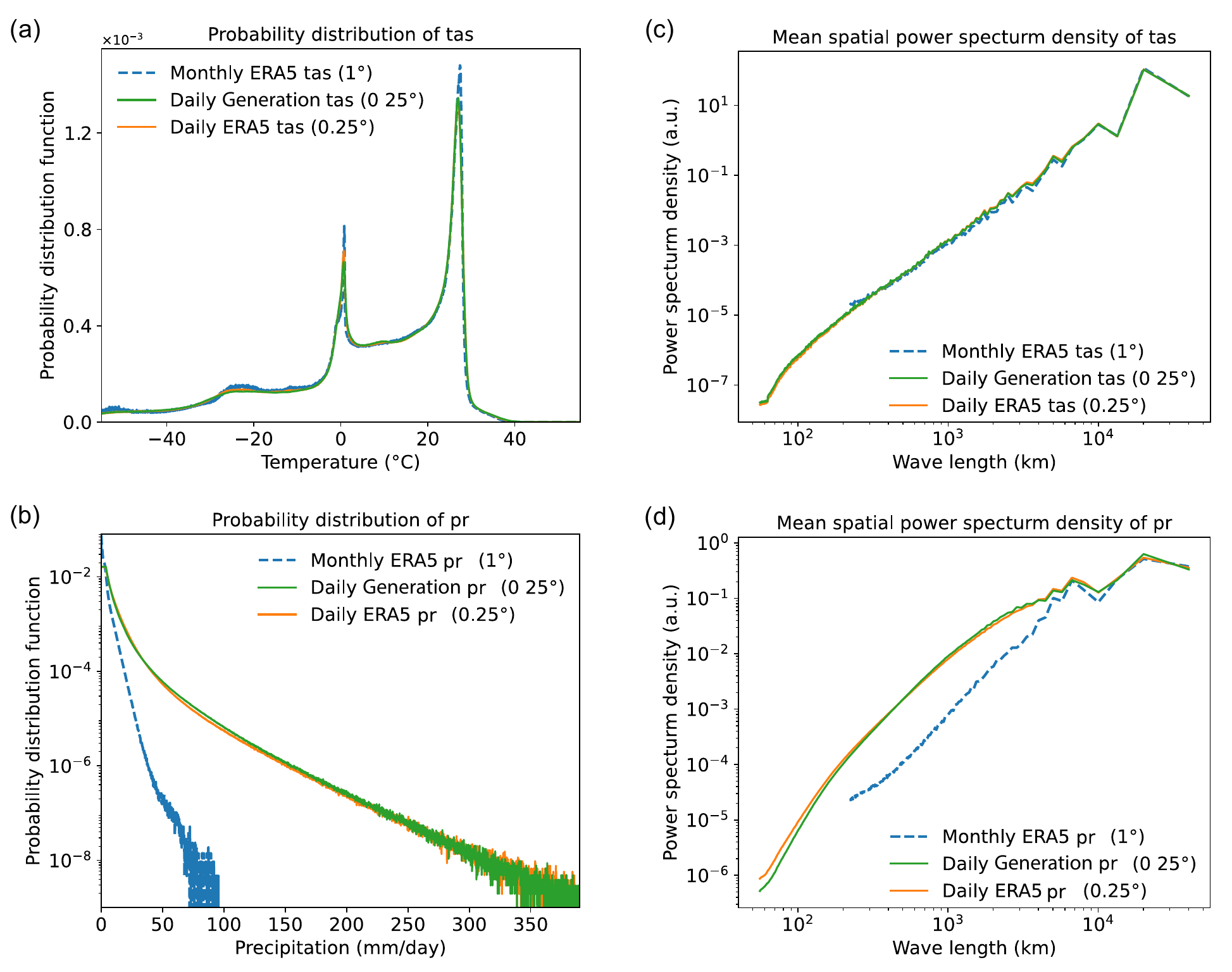}
\end{center}
\captionof{figure}{\textbf{Probability distribution and spatial power spectrum density of generated daily fields.} (a) Probability distribution functions of temperature (tas) from the monthly field, generated daily field, and ERA5 reference daily field. (b) Same as (a), but for precipitation (pr). (c) Mean spatial power spectral density of global temperature maps from the monthly field, generated daily field, and ERA5 reference daily field. (d) Same as (c), but for precipitation. All statistics are calculated over the testing period 2015--2023. }\label{fig5}\par\vspace{0.3cm}

\subsubsection{Probability distribution properties}
Across the full spatio-temporal domain, the empirical probability distributions of temperature and precipitation before and after downscaling are analyzed (Figs. 5a and 5b). Monthly mean fields, which represent averages over multiple days, exhibit distributional characteristics that differ substantially from those of daily fields. These monthly fields are used to guide the reconstruction of the probability distributions of daily variables. The results show that the generated daily temperature and precipitation are broadly consistent with the ERA5 daily reference. In particular, the heavy tail of the precipitation distribution is well reproduced in the generated daily fields (Fig. 5b).

In addition, the standard deviation of generated temperature and precipitation at each grid point characterizes the magnitude of local day-to-day fluctuations, reflecting the temporal variability of the underlying weather evolution. At the global scale, the generated fields produce consistent standard deviation with that in ERA5 reference (Figs. 6a and 6c), indicating that our method effectively captures the magnitude of day-to-day variability. However, a slight underestimation of variability is observed in some regions. For temperature, this occurs primarily over land in the mid- to high latitudes of both hemispheres, with a bias of approximately $-0.5\,^{\circ}\mathrm{C}$ (Fig. 6b). For precipitation, underestimation is mainly found in tropical regions and parts of the mid-latitude oceans, where variability is inherently large, corresponding to a relative deviation of about 5\%.

\begin{center}
\includegraphics[width=0.7\textwidth,keepaspectratio]{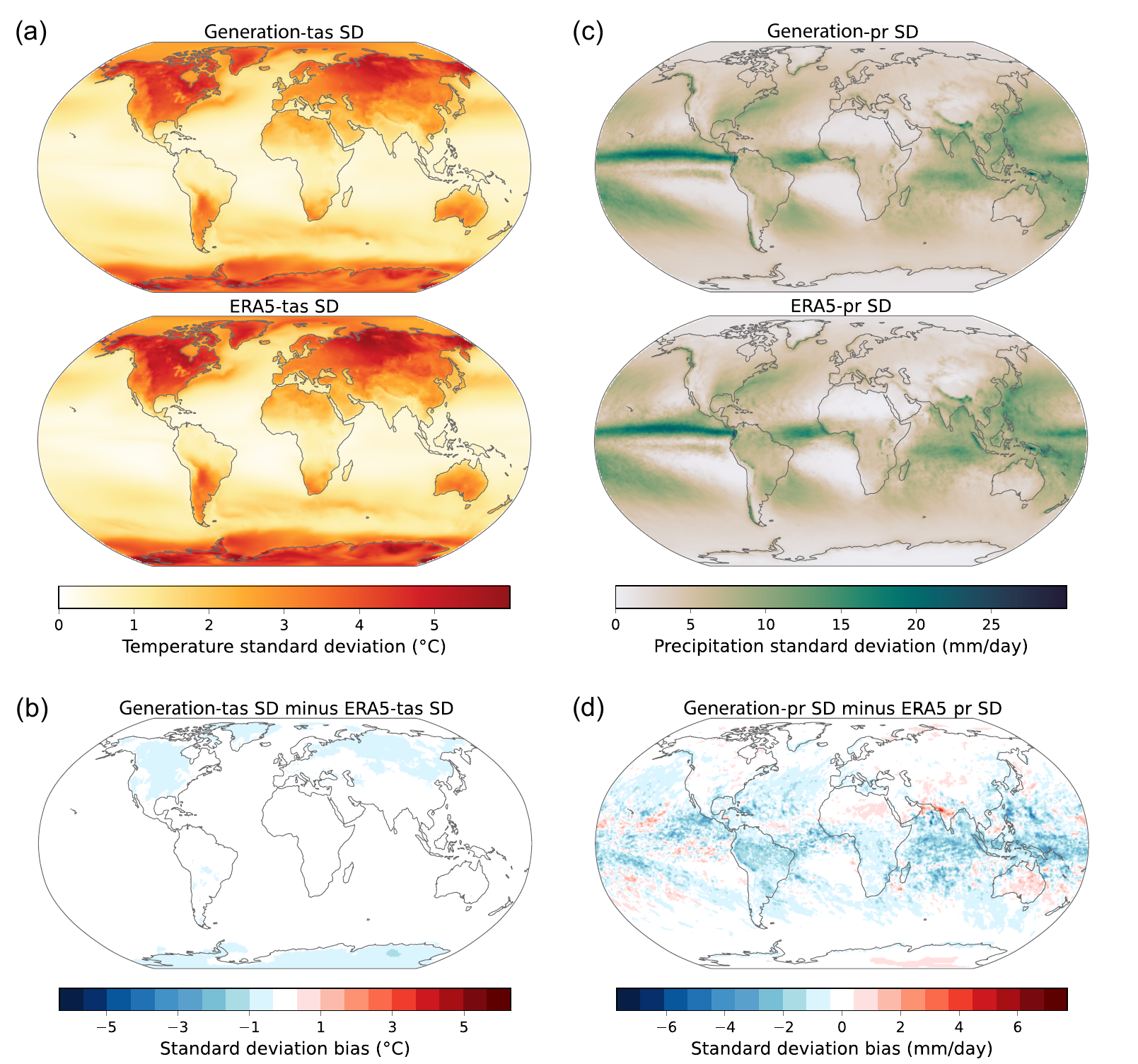}
\end{center}
\captionof{figure}{\textbf{Standard deviation for temperature and precipitation variability at individual grid cells of generated daily fields.} (a) Standard deviation (SD) of temperature (tas) from the generated daily field and ERA5 reference daily field. (b) Difference in standard deviation between the generated fields and ERA5. (c) Same as (a), but for precipitation (pr). (d) Same as (b), but for precipitation. All statistics are calculated over the testing period 2015--2023. }\label{fig6}\par\vspace{0.3cm}

\subsubsection{Spatial coherence and temporal persistence}

As a further examination on spatial coherence of the generated daily fields, we focus on the organization of spatial structures across grid points. This is quantified using the spatial power spectral density (PSD) of daily maps, with time-mean spectra shown in Figs. 5c and 5d. The original monthly fields exhibit a cut-off wavelength of approximately 222\,km, reflecting the $1^{\circ}$ spatial resolution (Nyquist limit). Monthly temperature fields display PSD characteristics similar to those of daily data, indicating comparable spatial coherence. In contrast, monthly-mean precipitation shows reduced power at high spatial frequencies, implying that downscaling must reconstruct fine-scale structures that are smoothed by temporal averaging.

The PSD of thegenerated daily fields is evaluated against that of the ERA5 reference. Across a wide range of wavelengths, the generation results closely match those of ERA5. In particular, the downscaling extends the effective wavelength range from 222\,km to 62\,km for both temperature and precipitation (Figs. 5c and 5d). The small-scale power of precipitation is well preserved (Fig. 5d), consistent with the visual improvements in spatial detail suggested by Figs. 2a and 2b. These results demonstrate that our model is able to reconstruct fine-scale spatial structures while maintaining the spatial coherence aligned with the observational reference.

\begin{center}
\includegraphics[width=0.8\textwidth,keepaspectratio]{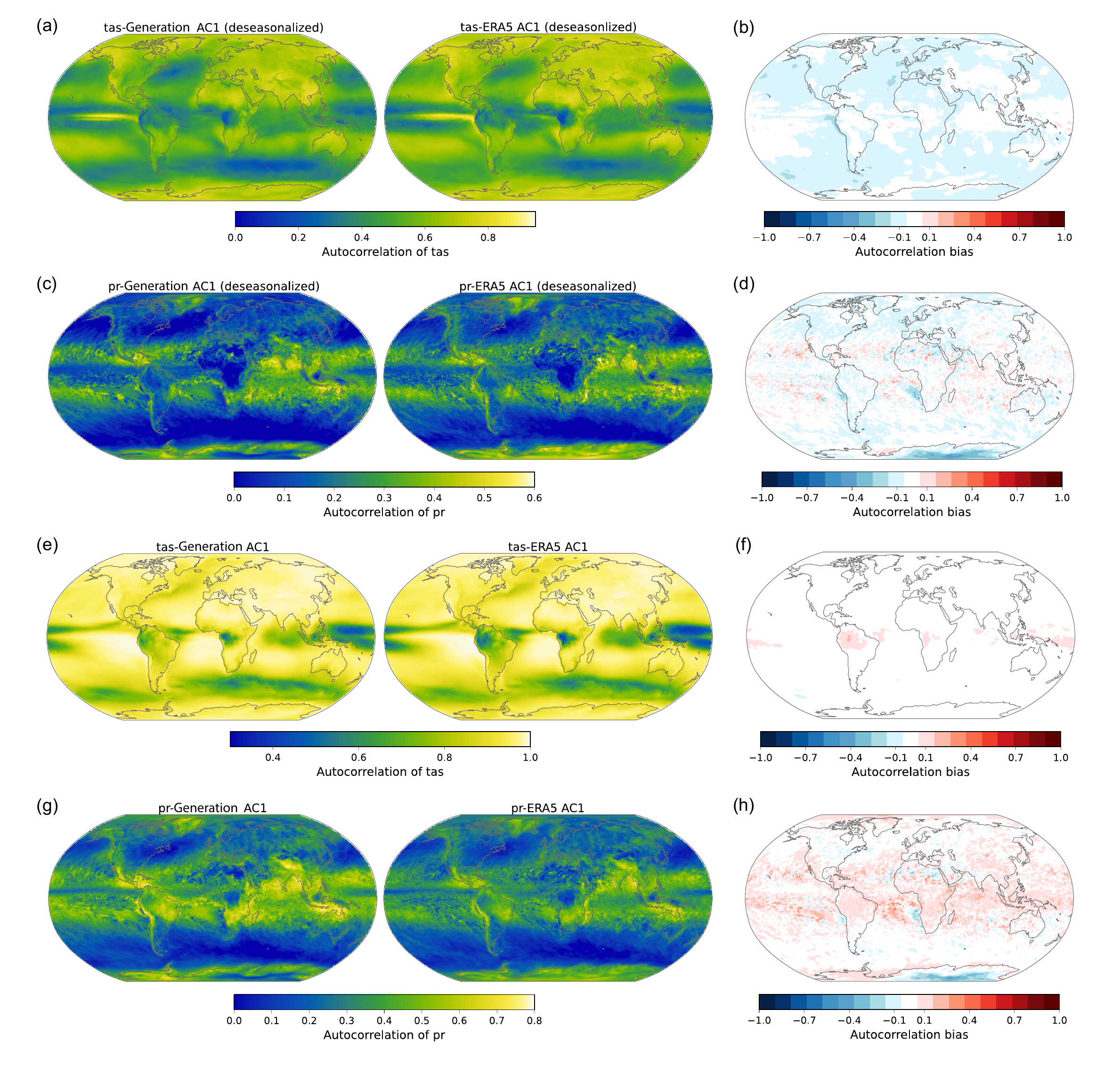}
\end{center}
\captionof{figure}{\textbf{Temporal persistence of temperature and precipitation at individual grid cells of generated daily fields.} (a) Lag-1 day autocorrelation (AC1) of temperature (tas) at each grid cell, computed from the generated field (left) and ERA5 reference daily field (right), respectively. The autocorrelation is estimated from time series deseasonalized by Seasonal-Trend decomposition using LOESS. (b) Difference in autocorrelation between generated fields and ERA5 shown in (a). (c)-(d) Same as (a)-(b), but for precipitation.  (e)-(h) Same as (a)-(d), but computed from the raw time series without deseasonalization. All statistics are calculated over the testing period 2015--2023. }\label{fig7}\par\vspace{0.3cm}

To assess whether the generated daily fields also exhibit temporal persistence consistent with the ERA5 reference, we compute the lag-1 autocorrelation at each grid point. This analysis is performed on both deseasonalized time series (Figs. 7a and 7c) and the original time series (Figs. 7e and 7g). In both cases, the model closely reproduces the spatial patterns of temporal persistence observed in ERA5. Deseasonalizing is conducted using seasonal trend decomposition based on LOESS (STL) \cite{cleveland1990stl, smith2023reliability}, removing both seasonal and long-term components and isolating high-frequency variability. This allows a direct assessment of the model's ability to capture day-to-day dynamics. Minor biases remain: autocorrelation is slightly underestimated (by less than 0.1) in some regions, while precipitation persistence is modestly overestimated (by less than 0.1) in parts of the subtropics. Relative to ERA5, the generated data exhibits slightly weaker persistence in mid-latitudes and slightly stronger persistence for precipitation in the tropics. In consideration of the underlying nonlinear temporal dependencies in precipitation, we in addition compute auto-mutual information~\cite{kraskov2004estimating,smith2015mutual} of precipitation for evaluation (Fig.~S1). The spatial distribution of this metric in generated daily precipitation closely matches that of ERA5 at the global scale, with only minor deviations in some mid-latitude regions and over Antarctica. The close agreement between generated temperature and precipitation and the ERA5 reference (Fig. 3) demonstrates that the model reliably captures the persistence of day-to-day variability at individual grid cells.

\subsubsection{Temperature-precipitation covariance}
Since the model jointly generates temperature and precipitation fields, covariance between the two variables is inherently expected in the generated outputs. To assess this, we compute the Pearson cross-correlation between temperature and precipitation at each grid cell. Based on the original time series, the estimated correlations show that the model closely matches ERA5 across all global grid points (Fig. 8a). The error distribution map (Fig. 8b) shows that the absolute deviations are all below 0.1, with no apparent differences between land and ocean regions. Correlation coefficients derived from the STL-deseasonalized data, which emphasize temperature-precipitation covariance in day-to-day variations, also demonstrate close agreement between the generated data and the ERA5 reference (Figs. 8c and 8d). We further use mutual information to quantify the dependence between daily temperature and precipitation anomalies at each grid cell. As shown in Fig.~S2, both the generated fields and ERA5 exhibit amplified mutual information over tropical regions. The discrepancies between the generated fields and ERA5 are generally small and are mainly confined to Antarctica. This demonstrates that the our model can effectively simulate and handle multiple variables jointly.

\begin{center}
\includegraphics[width=0.7\textwidth,keepaspectratio]{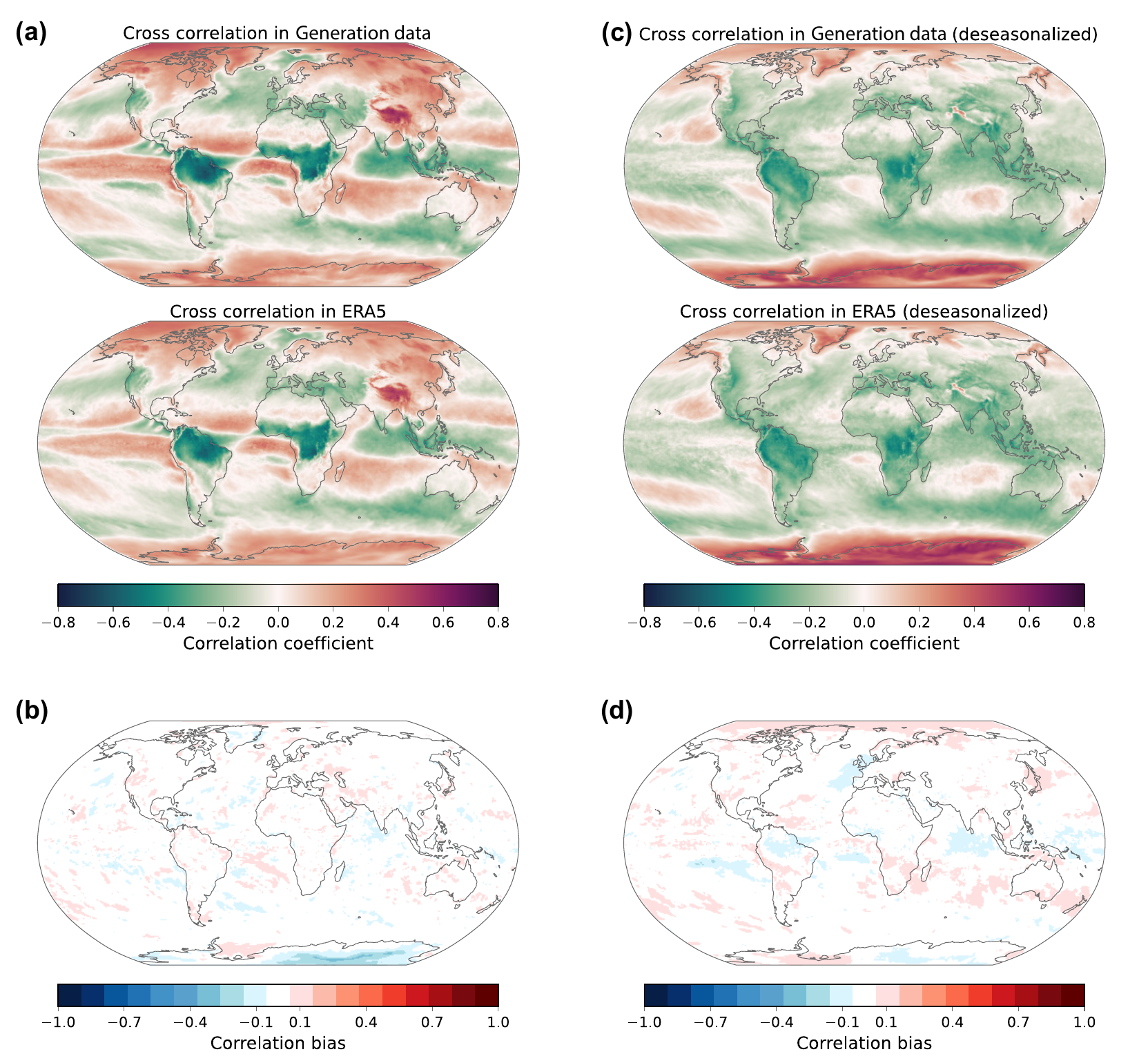}
\end{center}
\captionof{figure}{\textbf{Cross correlation between temperature and precipitation in generated daily fields.} (a) Pearson correlation coefficient between temperature and precipitation at each grid cell, derived from generated daily fields and ERA5 reference daily fields, respectively. Computation is from raw time series. (b) Difference in correlation coefficient between generated fields and ERA5 shown in (a). (c)-(d) Same as (a)-(b), but computed from the deseasonalized time series. All statistics are calculated over the testing period 2015--2023.}\label{fig8}\par\vspace{0.3cm}

\section{Summary and discussion}  
In this work, we present a generative machine learning method for downscaling monthly climate fields to daily ones while enhancing the spatial resolution from 1\textdegree{} to 0.25\textdegree{}, on a global scale. Our method is capable of jointly downscaling multivariate climate fields, as demonstrated for temperature and precipitation. The generated successive daily fields reproduce the spatiotemporal evolution patterns of temperature and precipitation, while maintaining close agreement with the ERA5 reference in key statistical properties, including probability distributions, extreme-value representation, spatiotemporal coherence, and inter-variable covariance. 

Our model builds a skillful foundation for flexibly generating climate fields at high spatiotemporal resultion from low-resolution input. By incorporating the guided generation technique, we showed that one can directly leverage a pre-trained model to efficiently generate daily fields under the guidance of monthly fields, without the need for additional fine-tuning or retraining. Moreover, our framework could be adapted to other tasks, such as downscaling weekly fields, regional fields or additional climate variables. Given that ERA5 represents a general circulation model with assimilated observational data, the application to ERA5 indicates that a similar spatiotemporal downscaling framework could be extended to other climate model datasets. However, despite its high quality, ERA5 still exhibits non-negligible biases when compared with high-resolution satellite and in situ observations \cite{lavers2022evaluation, chai2022evaluation, li2025land, sun2025fusion}, our diffusion model could in the future be trained or finetuned using satellite or in situ data \cite{sun2018review} to generate outputs that more closely reflect realistic conditions. 

Different from regional climate modeling (dynamical downscaling), our framework can be developed as an efficient and generalizable climate field generator that uses only the subset of desired climate variables to perform spatiotemporal downscaling of global fields. Such downscaled datasets exhibit statistical properties consistent with atmospheric dynamics and extreme-value representation. Looking forward, incorporating more climate variables may enable our generative machine learning model to capture more physical processes, such as feedbacks within the atmospheric circulation or ocean--atmosphere interactions, although this may also increase computational costs. Nevertheless, with the ongoing development and optimization of generative machine learning techniques, these challenges are likely to be better addressed in the future.

\section*{Code availability}
The diffusion model used in this study was implemented in Python~3.12 using PyTorch~2.2.0. All analyses were conducted in Python~3.12. The source code will be made publicly available upon publication via GitHub and Zenodo. Computation of spatial power spectrum was based on the torch-harmonic python library at https://github.com/NVIDIA/torch-harmonics.

\section*{Data availability}
The surface air temperature and total precipitation data were obtained from the Copernicus Climate Data Store for ERA5 reanalysis: \url{https://cds.climate.copernicus.eu/datasets/reanalysis-era5-single-levels}).The monthly ERA5 data was from:  \url{https://cds.climate.copernicus.eu/datasets/reanalysis-era5-single-levels-monthly-means}

\section*{Author contributions}
Y.H. and N.B. conceived this study. H.Y., N.B. and S.Y. designed this study. Y.H. and S.Y. performed programming. Y.H. and S.Y. improved the algorithms with contribution from M.A., and P.H.. Y.H. prepared the data and performed the numerical analysis. All authors contributed to discussion, interpretation and validation of the results. Y.H. performed the visualization. Y.H. wrote the manuscript with inputs from all authors.

\section*{Competing interests}
The authors declare that they have no competing interest.

\section*{Acknowledgements}
This is ClimTip contribution \# 162. The ClimTip project has received funding from the European Union's Horizon Europe research and innovation programme under grant agreement No. 101137601; The authors gratefully acknowledge the Ministry of Research, Science and Culture (MWFK) of Land Brandenburg for supporting this project by providing resources on the high performance computer system at the Potsdam Institute for Climate Impact Research. 
Y.H. acknowledges financial support by the Alexander von Humboldt Foundation. N.B. also acknowledges further support by the Volkswagen foundation, the Deutsche Forschungsgemeinschaft (DFG), the National Research Foundation of Korea (NRF): Climate Resilience under Zero-Emission Commitment Scenarios, BO 4455/2-1 (eBer-24-62950; RS-2024-00438471), the advanced Research and Invention Agency (ARIA), and the Past to Future (P2F) project, which has received funding from the European Union's Horizon Europe research and innovation programme under grant agreement No. 101184070

\printbibliography

\end{document}